\documentstyle[prd,aps]{revtex}
\begin{document}
\draft
\input epsf
 
 \title{\large\bf  STU Black Holes and String Triality}
%
\author{
\bf
Klaus Behrndt${}^a$,
Renata Kallosh${}^b$,
Joachim Rahmfeld${}^c$,
Marina Shmakova${}^{de}$, and
Wing Kai Wong${}^{b}$
\footnote{\tt
behrndt@qft2.physik.hu-berlin.de,
kallosh@renata.stanford.edu,
joachim@tam2000.tamu.edu,
shmakova@slac.stanford.edu,
wkwong@slac.stanford.edu}
}
%
\address{
${}^a$Humboldt-Universit\"at, Institut f\"ur Physik, Invalidenstra\ss
e 110,
10115 Berlin, Germany\\
${}^b$Physics Department, Stanford University, Stanford, CA
94305-4060\\
${}^c$Department of Physics, Texas A\&M University , College Station,
TX 77843\\
${}^d$University of Tennessee, Knoxville, TN 37996\\
${}^e$Stanford Linear Accelerator Center, Stanford University,
Stanford, CA 94309
}

\maketitle
\begin{abstract}
We found double-extreme black holes associated with the special
geometry of the
Calabi-Yau moduli space with the  prepotential
$F=STU$. The
area formula is  $STU$-moduli independent and has ${[SL(2, {\bf Z})]}^3$
symmetry in space of  charges.
The dual version of this theory without prepotential   treats the
dilaton $S$
 asymmetric versus  $T,U$-moduli. We display the dual
relation between new $(STU)$ black holes and stringy $(S|TU)$
black holes
using
particular $Sp(8, {\bf Z})$  transformation. The area formula of one
theory
equals the area formula of the dual theory when expressed in terms of
dual
charges.

We analyse the relation between $(STU)$ black holes to string
triality of
black holes:
$(S|TU)$, $(T|US)$, $(U|ST)$ solutions. In democratic $STU$-symmetric
version
we find that all three $S$ and $T$ and $U$ duality symmetries are
non-perturbative and mix electric and magnetic charges.

\end{abstract}
\pacs{PACS: 11.25.-w, 04.65.+e, 04.70.Dy  \hskip 1.5 cm  SU-ITP-96-37, HUB-EP-96/41,
~~hep-th/9608059 }


\section{Introduction}

The non-perturbative properties of the future fundamental theory
manifests
themselves in the duality properties of the area formulae  of the
supersymmetric
 black holes horizon.
The universal entropy-area formula  of supersymmetric black holes is
given by
the central charge extremized in the moduli space  $Z_{\rm fix}$ and
depends
only on quantized charges. The universal formula obtained by Ferrara
and one of
the authors \cite{FK}  is
$
S={A(p,q) \over 4}  =  \pi |Z_{\rm fix}|^\alpha$ with $  \alpha =2
\;(3/2)$ for
$ d=4\; (d=5)$.

This universal formula has various implementations in different
theories.  A
particularly rich class of area formulae may be expected to exist in
N=2
supersymmetric theories which are characterized by different choice
of the
holomorphic prepotential and/or symplectic sections.
A beautiful interplay between the geometry of special K\"ahler
manifolds
\cite{dWvP}-\cite{Cer}  and space-time geometry  of supersymmetric
black holes
has been discovered recently \cite{FKS}, \cite{FK}, \cite{ksw}.

In this paper we will find the 4d double-extreme black holes in a
class of N=2
theories
with the prepotential  $F=d_{ABC}\frac{X^AX^BX^C}{X^0}
$ \cite{CREM}.  These theories with real symmetric constant tensors
$d_{ABC}$
are related to geometry occurring in 5-dimensional supergravity
\cite{GST}
where
the term
$
\int d_{ABC} F^A \wedge F^B \wedge A^C
$
is present in the action. These theories are also related to the
special
geometry of Calabi-Yau moduli spaces where $d_{ABC}$ are the
intersection
numbers of the
Calabi-Yau manifold, $t^A={X^A\over X^0}$ are the moduli fields of
the K\"ahler
class \cite{FS}.
The theories of this class are also referred to as ``very special
geometry"
\cite{dWvP92} and ``real special geometry" \cite{BCDF}.

We will focus  mostly on $STU$-symmetric model
$F=\frac{X^1X^2X^3}{X^0},
$ and will find the moduli-independent
${[SL(2, {\bf Z})]}^3$
symmetric area formula. The
moduli of this theory are coordinates of the $\left
(\frac{SU(1,1)}{U(1)}\right)^3$ manifold. Duality symmetry of this
theory is
${[SL(2, {\bf Z})]}^3$.
The dual partners  of these black holes  (where one of the moduli,
e.g.   $S$ is
singled out and whose imaginary part
plays the role of string coupling)  are already known \cite{ksw}. The
moduli in
this version of the theory are coordinates of
$\frac{SU(1,1)}{U(1)}\times
\frac{SO(2,2)}{SO(2)\times SO(2)}$
 manifold. $S$-duality, or  $SL(2, {\bf Z})$  symmetry associated
with the $S$
moduli in string theory has a non-perturbative character, whereas $T$
and $U$
dualities, related to $SO(2,2)$ symmetry have perturbative character.
Perturbative symmetries of string theory do not mix electric and
magnetic
charges.
Stringy black holes treat one of the moduli on different footing than
others.
This is due to the fact that 11-dimensional supergravity has to be
reduced to
$d=10$ first and this makes 11-th component  of the metric or the
dilaton,
special. If however, we are looking for exact non-perturbative
solution of
11-dimensional supergravity, we may expect some solutions where the
radius of
11-th, 6-th and 5-th dimensions are all on equal footing. These are
our $STU$
black holes. They may be related to M, F, Y or whatever fundamental
theory which is not the conventional theory of strings.
To establish the  relation between new  $(STU)$-symmetric black
holes and
their dual   $(S|  TU)$ stringy partners is our main goal. In string
triality
picture \cite{Triality} the role of $S$ may be replaced by $T$ or $U$
but still
there is one moduli different from the others and only one duality
symmetry is
non-perturbative whereas the other two are perturbative. We will find
that all
three $S$ and $T$ and $U$ duality symmetries in ``democratic" black hole
solutions are
non-perturbative. This  is not too surprising: black holes are
non-perturbative
objects!

We will find that the area of the horizon in $(STU)$-symmetric theory
equals the area of the horizon of the $(S|TU)$ dual theory
\begin{equation}
A^{(STU)}(p,q)= A^{(S | TU)} (\hat p, \hat q) \ ,
\label{area}\end{equation}
where
the charges are related by particular $Sp(8, {\bf Z})$ duality
transformation.
This transformation has been found in \cite{Cer} and relates the
symplectic
sections and charges in two theories.

The duality symmetry of this area formula is of an unusual form. The
typical
situation studied before was that  the area as a function of charges
was
invariant
under duality transformation.
\begin{equation}
A(p,q) = A (\hat p, \hat q) \ .
\end{equation}
For example, U-duality \footnote{U-duality in the context of
$E_7$-symmetry
should not be confused with
$U$-duality in the context of $SL(2,{\bf Z})$ symmetry related to
$U$-moduli.
Unfortunately, these two different dualities carry the same name in
the current
literature. } invariant area formula is given by the quartic Cartan
invariant
of $E_{7}$ in d=4,  \cite{KK},  \cite{FK}, \cite{CH}, where the
$2\times 28$
unhatted
$(p,q)$ are charges before duality transformation and  $2\times 28$
hatted
$(\hat p,  \hat q)$ are charges after  $E_{7}$ transformation.  This
duality
transformation was a property of one specific theory:
in this case, for example, N=8 supergravity in d=4. The equations of
motion of
this theory have  hidden symmetry  and it manifests itself in $E_{7}$
invariance of the area formula of the black holes of this theory with
1/8 of
supersymmetry unbroken.
\begin{equation}
A^{N=8}(p,q)= A^{N=8} (\hat p, \hat q) = 4\pi \sqrt {J(p,q)}= 4\pi
\sqrt
{J(\hat p, \hat q)} \ .
\end{equation}
The new phenomenon which we observe here by studying the black holes
in the
framework of special geometry  is the following.  Black holes in  two
versions
of the theory related by symplectic transformation
have two different area formulae, when the area of the original
version is
expressed in terms of charges of original theory and the area of the
transformed (dual) theory is expressed as a function of charges of
dual theory.
 However, these two area formulae are related as in eq.
(\ref{area}). If one
has the area in one theory and the transformation which defines the
dual theory
is known, the area can be found using (\ref{area}).
The reason for the area formulae to be different is that they carry
different 
symmetries: $\left (\frac{SU(1,1)}{U(1)}\right)^3$ in one case and
$\frac{SU(1,1)}{U(1)}\times \frac{SO(2,2)}{SO(2)\times SO(2)}$ in the
other
case.

The paper is organized as follows. In Sec. 2 we discuss the basic
equations
\cite{FK} defining the double-extreme black holes of N=2 theory \cite{ksw}  and
the values
of moduli as functions of charges. We refer to these equations as
``stabilization equations".  The main property of these equations
relevant to
present investigation is that they are symplectic covariant.
Therefore once the
 solution for moduli in terms of charges is known in one version of
the theory,
the dual solution can be found by applying the symplectic
transformation to the
known solution.  We explain this for the case of  ST[2,n] manifold,
${SU(1,1)\over U(1)} \times {SO(2,n)\over SO(2)\times SO(n)}$
symmetric
theory which does not admit a prepotential and the dual version of it
which
admits the prepotential. We also explain that in both theories one
has an
option of solving for double-extreme black holes directly in each
version,
without using the information on the solution in the dual theory.
Having
obtained these two sets of double-extreme black holes
one can check that the solutions are actually connected by symplectic
transformation. Or one could use the solution available on one side
and perform
a relevant symplectic transformation to get the double-extreme black
holes of
the dual theory and the mass-area formula in terms of dual charges.
One can
verify that the transformed solution indeed solves the equations of
the dual
theory. In Sec. 3 we proceed with solving ``stabilization equations"
on the
prepotential side and we consider the  prepotential  $F=\frac{X^1X^2X^3}{X^0}$.
 We derive the new mass-area formula for this theory. In
Sec. 4  we
show that alternative derivation of $STU$ symmetric double-extreme
black holes
is possible: via symplectic transformation from  the dual version of
the theory
without the prepotential, i.e. from the theory where $S$ is not
symmetric with
$T,U$.
In Sec. 5 these two sets of double-extreme black holes are studied
from the
perspective of string triality and the difference between the new
 and stringy  black holes
solutions is explained.
In the Outlook  we point out the implication of our new d=4 area
formulae for
Calabi-Yau moduli space and the corresponding d=5 area formulae. We
also
comment on  string loop corrections and their possible effect on
supersymmetric
black holes and vice versa.

\section{Symplectic covariance of ``stabilization equations"}

Stabilization equations for   $n_v$ complex   moduli of
supersymmetric black
holes in N=2 theory near the
horizon have the following form \cite{FK}
\begin{equation}
\left (\matrix{
p^\Lambda\cr
q_\Lambda\cr
}\right )={ \rm Re} \left (\matrix{
2i \bar Z L^\Lambda\cr
2i \bar Z M_\Lambda\cr
}\right )  ,
\label{stab}
\end{equation}
where
the central charge \cite{Cer}
\begin{equation}
Z(z, \bar z, q,p) = e^{K(z, \bar z)\over 2}
(X^\Lambda(z)  q_\Lambda - F_\Lambda(z) \, p^\Lambda)= (L^\Lambda
q_\Lambda -
M_\Lambda p^\Lambda)  
\label{central}\end{equation}
depends on moduli and on $2n_v+2 $
conserved charges
($p^\Lambda,q_\Lambda $).  ($L^\Lambda, M_\Lambda $) are covariantly
holomorphic sections
depending on moduli. For double-extreme black holes \cite{ksw} with
frozen
moduli these equations implicitly define the frozen moduli as
functions of
charges.

Symplectic transformation acts on charges as well as on sections
\begin{equation}
\left (\matrix{
\hat p\cr
\hat q \cr
}\right ) = \pmatrix{
A & B \cr
C & D \cr
} \left (\matrix{
p\cr
q\cr
}\right ) , \hskip 3 cm \left (\matrix{
\hat X\cr
\hat F \cr
}\right ) = \pmatrix{
A & B \cr
C & D \cr
} \left (\matrix{
X\cr
F\cr
}\right ),
\label{symplec}\end{equation}
and provides the relation between the dual versions of the theory.
Here
\begin{equation}
\pmatrix{
A & B \cr
C & D \cr
} \in Sp(2n_v +2, {\bf Z})\ . \end{equation}
 Stabilization equations are covariant under  symplectic
transformations
\begin{equation}
\left (\matrix{
\hat p^\Lambda\cr
\hat q_\Lambda\cr
}\right )={ \rm Re} \left (\matrix{
2i \bar Z \hat L^\Lambda\cr
2i \bar Z \hat M_\Lambda\cr
}\right ) \ ,
\label{stab1}
\end{equation}
since the central charge is invariant.
 \begin{equation}
Z =  (L^\Lambda
q_\Lambda -
M_\Lambda p^\Lambda)  =  (\hat L^\Lambda
\hat q_\Lambda -
\hat M_\Lambda \hat p^\Lambda)\ .
\label{central2}\end{equation}
and the sections are covariant.

We are interested in dual relation between black holes of two
theories. The
first one
 can
be defined
in terms of a prepotential  \footnote{We are using here notation of
\cite{ABCAFF}.}
\begin{equation}
F= {1\over 2} d_{ABC} t^A  t^B  t^C = S\,  \eta_{\alpha  \beta } \, t^\alpha  t^\beta \ ,
\qquad X^0
=1 \ ,
\end{equation}
where
\begin{equation}
t^1=S, \qquad d_{ABC}= \left (\matrix{
d_{1 \alpha  \beta}= \eta_{\alpha  \beta} \cr
0 \; \rm{otherwise} \cr
}\right ) , \hskip 2 cm A,B,C = 1,2, \dots , n+1 \ ,
\end{equation}
and
\begin{equation}
 \eta_{\alpha  \beta} = {\rm diag} (+,-,- \dots , -) , ~~~\quad  \alpha,   \beta = 2, \dots ,
n+1.
\end{equation}
This prepotential corresponds to the product manifold
$\frac{SU(1,1)}{U(1)}\times \frac{SO(2,n)}{SO(2)\times SO(n)}$. The
$\frac{SU(1,1)}{U(1)}$ coordinate is the axion-dilaton field $S$. The
remaining
$n$ complex moduli $t^i$ are special coordinates  of the
$\frac{SO(2,n)}{SO(2)\times SO(n)}$ manifold. In particular, when
$n=2$ we have
\begin{equation}
F= {1\over 2} d_{ABC} t^A  t^B  t^C = {1\over  2} S\left[ (t^2)^2-
(t^3)^2\right] \ .
\end{equation}
This theory is defined by 3 complex moduli and 4 gauge groups and the
corresponding manifold is $\frac{SU(1,1)}{U(1)}\times
\frac{SO(2,2)}{SO(2)\times SO(2)}$.

If we introduce the notation
\begin{equation}
t^2 \equiv {1\over \sqrt 2} (T+U) \ , \qquad t^3 \equiv {1\over \sqrt
2} (T- U)
 \ ,
\end{equation}
the prepotential becomes
\begin{equation}
F =  STU \ .
\label{prep}\end{equation}
This theory has the  symmetry of the manifold $\left
(\frac{SU(1,1)}{U(1)}\right)^3$, which corresponds to the embedding
of $\left
(\frac{SU(1,1)}{U(1)}\right)^2$ into $\frac{SO(2,2)}{SO(2)\times
SO(2)}$.
The $STU$ symmetric theory with  the cubic
holomorphic prepotential (\ref{prep}) is associated with particular
Calabi-Yau
moduli space.
It  is related to the dual version of the theory
via symplectic transformation  \cite{Cer}. We will study this
relation in the
context of black holes
in Sec. 4.

There are two possibilities to find double-extreme black holes in the
theory
with the prepotential. Either directly solve the stabilization
equations or
perform the dual rotation from the known solution. In the next
section we will
first find the black holes directly in $STU$-symmetric model.

\section{Double-extreme black holes in STU-model }
STU-model \cite{STU} is  described by the prepotential:
\begin{equation}
F(X)={d_{ijk}X^iX^jX^k \over { X^0}}\ , \qquad  i,j,k=1,2,3,
\end{equation}
For our case
\begin{equation}
F(X)={X^1X^2X^3 \over X^0} \ .
\end{equation}
Holomorphic section  determined by that prepotential has a form:
 $(X^\Lambda, F_{\Lambda})$ with $ F_\Lambda={\partial F \over
\partial
X^\Lambda } $ and $\Lambda =(0, i=1,2,3)$.
Special coordinates $ z^i $ are determined by
\begin{equation}
z^i={X^i \over X^0} \ ,  \qquad  X^0=1 \ .
\end{equation}
Corresponding K\"ahler potential  is
\begin{equation}\label{kp}
K=-\log\big ( -id_{ijk}(z-\bar z)^i(z-\bar z)^j(z-\bar z)^k \big )\ .
\end{equation}
and we will use also
\begin{equation}
z^1=S, \quad z^2=T\ ,\quad z^3=U\ .
\end{equation}

In terms of special coordinates the holomorphic sections are given
by:
\begin{equation}
X^\Lambda =  \left(\matrix { 1 \cr
                     z^1 \cr
                     z^2 \cr
                     z^3 \cr} \right ), \qquad
F_\Lambda =  \left(\matrix { {-z^1z^2z^3} \cr
                     z^2z^3 \cr
                     z^1z^3 \cr
                     z^1z^2 \cr}\right ).
\end{equation}

 The stabilization equations are:
\begin{eqnarray}
p^\Lambda =ie^{K/2}\bigl( {\bar Z}X^\Lambda -Z{\bar X}^\Lambda \bigr
)\ ,\\
q_\Lambda =ie^{K/2}\bigl( {\bar Z}F_\Lambda -Z{\bar F}_\Lambda \bigr
)\ .
\end{eqnarray}

We can eliminate  $ \bar Z $ from these equations so that:
 \begin{equation}\label{stabnew}
X^\Lambda q_\Sigma -p^\Lambda F  _\Sigma  =ie^{K/2}Z\bigl( {\bar
X}^\Lambda
F_\Sigma -{X}^\Lambda {\bar F}_\Sigma \bigr )\ .
\label{matrix}\end{equation}

This is the matrix equation we used in \cite{ksw} to solve for the
solution of
frozen moduli.
In what follows we will  solve for $z^1$ as a function of
charges.  The solution for $z^2$ and $z^3$ can be obtained in an
analogous
way as a result of symmetry between the three moduli.

Here are the components ($(\Lambda, \Sigma) = (1,0), (0,1), (1,1),
(2,3)
\text{\ and\ } (3,2)$ respectively)
from the matrix equation (\ref{matrix}) we  need for the derivation
of $z^1$:
\begin{eqnarray}
q_0 +p^1z^2z^3 &=&ie^{K/2}Z(\bar z^1 \bar z^2\bar z^3 -{\bar z^1}
z^2z^3)
\label{10}\ ,\\
q_1 -p^0 z^2z^3 &=&ie^{K/2}Z( z^2z^3- \bar z^2 \bar z^3) \label{01}\ ,\\
q_1 z^1-p^1 z^2z^3 &=&ie^{K/2}Z(\bar z^1 z^2z^3 - z^1 \bar z^2 \bar
z^3)
\label{11}\ ,\\
q_3 z^2-p^2 z^1z^2 &=&ie^{K/2}Z(\bar z^2 z^2z^1-z^2 \bar z^2 \bar
z^1)\label{23}\ ,\\
q_2 z^3-p^3 z^1z^3 &=&ie^{K/2}Z(\bar z^3 z^3z^1-z^3  \bar z^3 \bar
z^1)\ .\label{32}
\end{eqnarray}

Using (\ref{10}) and (\ref{01}) we can eliminate the factor $ie^{K/2}
Z$ and
obtain
\begin{equation}
 z^2z^3={q_1 \bar z^1 +q_0 \over p^0 \bar z^1 - p^1 } \label{z2z3}\ .
\end{equation}

Using (\ref{11}, \ref{z2z3})  we can obtain a simple  formula for $i
e^{K/2} Z$
\begin{equation}
\label{B1}
 ie^{K/2}Z={( p^0 z^1 - p^1) \over (\bar z^1 - z^1) }\ .
\end{equation}

Substituting (\ref{B1})  into (\ref{23}) and (\ref{32})  respectively
we can
express $z^2$  and $z^3$ in  terms of $z^1$ and the charges only:

\begin{equation}
\bar z^2={( p^2 z^1 - q_3) \over (p^0 z^1 - p^1)}\ , \quad \text{ and }
\quad
\bar z^3={( p^3 z^1 - q_2) \over (p^0 z^1 - p^1)}\ .
\end{equation}

Finally,  using (\ref{z2z3}) and the above equations to eliminate
$z^2$ and
$z^3$
we are getting a quadratic equation for $z^1$

\begin{equation}
{(z^1)}^2+{((p\cdot q)-2p^1q_1)\over
(p^0q_1-p^2p^3)}z^1-{(p^1q_0+q_1q_2)\over
(p^0q_1-p^2p^3)}=0 \ ,
\end{equation}

where
\begin{equation}
(p\cdot q)= (p^0q_0) +(p^1q_1)+(p^2q_2)+(p^3q_3) \equiv  p^\Lambda
q_{\Lambda}\ ,
\end{equation}
and the solution for $z^1$ moduli is 
\begin{equation}
z^1={((p\cdot q)-2p^1q_1)\over 2(p^3p^2-p^0q_1)}\mp i{\sqrt{W}\over
2(p^3p^2 -
p^0q_1)} \ ,
\end{equation}
where
\begin{equation} \label{ww}
 W(p^\Lambda ,q_\Lambda) =-{(p\cdot q)}^2+4\bigl (
(p^1q_1)(p^2q_2)+(p^1q_1)(p^3q_3)+(p^3q_3)(p^2q_2)\bigr )\\
 - 4 p^0 q_1 q_2 q_3 + 4q_0 p^1 p^2 p^3 \ .
\end{equation}
The function $ W(p^\Lambda ,q_\Lambda)$ is symmetric under
transformations: $
p^1\leftrightarrow p^2 \leftrightarrow p^3 $
and  $ q_1\leftrightarrow q_2 \leftrightarrow q_3. $
Finally solution for all three complex  modulus are:
\begin{equation}
z^i={((p\cdot q)- 2p^iq_i )\mp i \sqrt{W}\over  2(3 d_{ijk} p^j p^k-
p^0q_i)} \ .
\label{sol}\end{equation}
There is no summation over $i$ in $p^iq_i$. For the solution to be
consistent
we have to require $ W>0 $,
otherwise the moduli are real and the K\"ahler potential is not defined

At this point the choice of signs in the imaginary part of the moduli
is ambiguous. However, to preserve the obvious exchange symmetries,
we want to
choose common signs for all. In fact it turns out that only the "$-$"
is
consistent as we shall see.

With these expressions for the $z^i$ the K\"ahler potential
eq. (\ref{kp}) is
easily computed.
We find
\begin{equation}\label {kp1}
e^{-K}=\pm \frac{W^{3/2}}{\omega_1 \omega_2 \omega_3}\ ,
\end{equation}
where
\begin{equation}
\omega_i=(3 d_{ijk}p^jp^k-p^0q_i) \ .
\end{equation}

It is also useful to calculate the product of three $\omega_i$'s
which appears
to be positive:
\begin{equation}
 \omega_1 \omega_2 \omega_3 = {1\over 4}  \left( (p^0)^2W
+[ 2p^1p^2p^3 - p^0 (p\cdot q)]^2\right) > 0
\end{equation}
with $\Lambda=0,1,2,3$,

For  the  K\"ahler potential $e^{-K}$ to be positive we have to pick
up only
one choice of sign for each imaginary part of the special coordinates
in eq.
(\ref{sol}), it has to be negative:
\begin{equation}
z^i={((p\cdot q)- 2p^iq_i )- i \sqrt{W}\over  2(3 d_{ijk} p^j p^k-
p^0q_i)}  
\qquad \Longrightarrow
\qquad
e^{-K}= \frac{W^{3/2}}{\omega_1 \omega_2 \omega_3} >0\ .
\label{STUsolution1}\end{equation}
We can proceed now with the calculation of the central charge to find
the black
hole mass, which for double-extreme black holes is proportional to
the area of
the black hole horizon. We find that
 \begin{equation}
e^{K } Z \bar Z =    \frac{(p^0)^2W +[2p^1p^2p^3 - p^0 ( p\cdot q)]^2}{4W} \ .
\end{equation}

We deduce for the mass/area
\begin{equation}Z\bar Z=M^2=  \frac{W^{3/2}}{\omega_1\omega_2\omega_3}\frac{(p^0)^2W
+[2p^1p^2p^3 - p^0 ( p\cdot q)]^2}{4W} \ ,
\label{area1}
\end{equation}
which finally gives the beautiful result
\begin{equation}
 Z \bar Z = M^2=   {A\over 4\pi} =  \left( W(p^\Lambda
,q_\Lambda)\right)
^{1/2} \ .
\end{equation}
This is a very nice and simple expression for the area which relies
on the
fact that the nominator of the second expression
in (\ref{area1}) and the product of the   $\omega_i$ cancel.
 Thus we have completely described the double-extreme
black
holes solutions with frozen moduli in the  $STU$ symmetric theory.
The geometry
is that of extreme Reissner- Nordstr\"om type with the mass/area  formula,
as
function of quantized charges given in eq. (\ref{ww}).
\begin{equation}
ds^2 =\left ( (1+  {[W (p,q)]^{1/4}\over r}\right)^{-2} dt^2  - \left ( (1+ {[W
(p,q)]^{1/4}\over r}\right)^{2} d\vec x^2 \ .
\end{equation}
It is instructive to remind that our mass/area formula has also a nice
symplectic invariant form, as explained in \cite{FK}, \cite{ksw}.
\begin{eqnarray}
M^2
&=& -\frac{1}{2}
\left(
p^\Lambda,
q_\Lambda \right)
\pmatrix{
({\text {Im\,}} {\cal N} +{\text {Re\,}} {\cal N} {{\text {Im\,}} {\cal
N}}^{-1}{\text {Re\,}} {\cal N})_{\Lambda\Sigma}   &{(-{\text
{Re\,}} {\cal N}
{{\text {Im\,}}
{\cal N}}^{-1})_\Lambda}^\Sigma \cr
({-{\text {Im\,}}  {\cal N}}^{-1} {\text {Re\,}} {\cal N})^\Lambda_\Sigma
 &( {{\text {Im\,}} {\cal N}}^{-1})^{\Lambda\Sigma} \cr
}_{\rm fix}
\left (\matrix{
p^\Sigma\cr
q_\Sigma\cr
}\right )
\nonumber\\
\nonumber\\
&=&
-\frac{1}{2}
\left(
p^\Lambda,
q_\Lambda \right)
\pmatrix{
({\text {Im\,}} F +{\text {Re\,}} {F} {{\text {Im\,}} F}^{-1}{\text {Re\,}}
F)_{\Lambda\Sigma}   &{(-{\text
{Re\,}}F
{{\text {Im\,}}
F}^{-1})_\Lambda}^\Sigma \cr
({-{\text {Im\,}} F}^{-1} {\text {Re\,}} F)^\Lambda_\Sigma
 &( {{\text {Im\,}} F}^{-1})^{\Lambda\Sigma} \cr
}_{\rm fix}
\left (\matrix{
p^\Sigma\cr
q_\Sigma\cr
}\right )\ \ ,
\label{mass2}\end{eqnarray}
where the period matrix ${\cal N}$ as well as the second derivative of the
prepotential $F$ are  functions of moduli which at the fixed point near the
black hole horizon become functions of charges, as defined in
eq. (\ref{STUsolution1}).

 If we parametrize
 all 3 moduli in terms of axion-dilaton fields
\begin{equation}
z^i=a^i-ie^{-\eta_i} \end{equation}
where
\begin{equation}
a^i= {((p\cdot q)-2p^iq_i) \over 2 \omega_i}\ , \qquad e^{-\eta_i}=
{\sqrt{W}\over
2 \omega_i}\ ,
\end{equation}
K\'ahler potential is
\begin{equation}\label{kpNew}
e^{-K}=  -8\, {\rm Im} S\   {\rm Im} T\    {\rm Im} U  = 8
e^{-\eta_1}e^{-\eta_2}e^{-\eta_3}\ .
\end{equation}
 This parametrization is possible under the condition that all three
combination
of charges
are positive, \begin{equation}
\omega_i >0 \ .
\end{equation}

\section{Dual rotation of double-extreme black holes}

The  double-extreme
black holes
for this model  without the prepotential  for the general case of
arbitrary $n$ as well as for  $n=2$  have been found before
\cite{ksw}. The
resume of this black hole
for $n=2$ is the following. Solution is defined in terms of  4
magnetic and 4
electric and charges $(\hat p^\Lambda , \hat q_\Lambda) $  and
$\Lambda
=0,1,2,3$.
The frozen moduli
are given by \footnote{ We are choosing the negative sign for the
imaginary
part of $S$ here for the sake of the dual rotation to the
prepotential version,
using the symplectic matrix (\ref{ourcase}). }
\begin{equation}
S= {\hat p \cdot \hat q   - i \left (\hat p^2 \hat q^2 - (\hat p
\cdot \hat
q)^2\right )^{1/2} \over \hat p^2} \ ,
\end{equation}

\begin{equation}
T= {\hat X^3 - \hat X^1 \over \hat X^0 - \hat X^2} = {\bar S (\hat
p^3 - \hat
p^1) - (\hat q^3 - \hat q^1) \over \bar S (\hat p^0 - \hat p^2) -
(\hat q^0 -
\hat q^2) } \ ,
\end{equation}

\begin{equation}
U= {- \hat X^3 - \hat X^1 \over \hat X^0 - \hat X^2} = {\bar S (-\hat
p^3 -
\hat p^1) - (-\hat q^3 - \hat q^1) \over \bar S (\hat p^0 - \hat p^2)
- (\hat
q^0 - \hat q^2) } \ ,
\end{equation}
and the mass/area formula is
\begin{equation}
Z \bar Z = M^2 = {A\over 4\pi} = \left (\hat p^2 \hat q^2 - (\hat p
\cdot \hat
q)^2\right )^{1/2} \ .
\end{equation}
The symplectic transformation between the theory without the
prepotential
(``hatted" version) to the one with the prepotential (``unhatted"
version)
is\cite{Cer}:
\begin{equation} \label{spmatrix}
Sp(8, {\bf Z}) \ni
\pmatrix{
A & B \cr
C & D \cr
}= \pmatrix{
A & B \cr
-B & A \cr
}
\end{equation}
with
\begin{equation}
A= {1\over \sqrt 2} \left (\matrix{
1 & 0 & 0 & 0 \cr
0 & 0 & -1 & -1 \cr
-1 & 0 & 0 & 0 \cr
0 & 0 & 1 & -1 \cr
}\right ), \qquad B = {1\over \sqrt 2} \left (\matrix{
0 & -1 & 0 & 0 \cr
0 & 0 & 0 & 0 \cr
0 & -1 & 0 & 0 \cr
0 & 0 & 0 & 0 \cr
}\right ).
\label{ourcase}\end{equation}

Starting with the prepotential  $F =  STU$
in terms of special coordinates we have the holomorphic section:
\begin{equation}
X^\Lambda =  \left(\matrix { 1 \cr
                    S \cr
                     T \cr
                     U \cr} \right ), \qquad
F_\Lambda =  \left(\matrix { {-STU} \cr
                     TU \cr
                     SU \cr
                     ST \cr}\right ).
\end{equation}
After symplectic transformation defined in (\ref{symplec}),
(\ref{ourcase}) we
get for hatted sections:
\begin{equation}
\hat X^\Lambda =  {1\over \sqrt 2} \left(\matrix { 1-TU \cr
                     -(T+U) \cr
                     -(1+TU) \cr
                     T-U \cr} \right ), \qquad
\hat F_\Lambda =   \left(\matrix { S \hat X^0 \cr
                     S \hat X^1 \cr
                     -S \hat X^2 \cr
                    -S \hat X^3 \cr}\right ) = S\eta_{\Lambda \Sigma}
\hat X^\Sigma \ ,
\end{equation}
where the  metric  is $\eta_{\Lambda \Sigma}  = (+ + --)$. This
theory does not
admit the
prepotential \cite{Cer}.

We can now relate the known results of the version without a
prepotential
(for which we use variables with a hat) to the ones obtained here.
{}From eqs. (\ref{symplec}),  (\ref{ourcase}) we find the
transformation
between $\hat p, \hat q$ and
$p, q$ to be:
\begin{equation}
\left [\matrix{
\hat p^\Lambda\cr
\hat q_\Lambda\cr
}\right ]= \frac{1}{\sqrt{2}}
 \left [\matrix{
p^0-q_1\cr
-p^2-p^3\cr
-p^0-q_1\cr
p^2-p^3\cr
p^1+q_0\cr
-q_2-q_3\cr
p^1-q_0\cr
q_2-q_3\cr
}\right ]
\label{charges}\end{equation}
This transformation gives us the relations:
\begin{eqnarray}
&&\hat p^2 =  2(p^2p^3- p^0q_1) = 2\omega_1 \ ,\\
&&\hat p \cdot \hat q   =  p\cdot q - 2 p^1 q_1 \ ,
\end{eqnarray}
hence we find
\begin{equation}
 S = z_1 \ ,
\end{equation}
where $ S$ is the first moduli  field of the version without the prepotential,
and (with a little more work) we have
\begin{equation}
\hat p^2 \hat q^2 -(\hat p \cdot \hat q)^2 = W \ .
\end{equation}
For $T$ and $U$ we get, using the relation between charges
(\ref{charges})

\begin{equation}
T=  {\bar S (\hat p^3 - \hat p^1) - (\hat q^3 - \hat q^1) \over \bar
S (\hat
p^0 - \hat p^2) - (\hat q^0 - \hat q^2) } = {\bar S  p^2 - q_3 \over
\bar S
p^0 - p^1 }=z_2 \ ,
\end{equation}

\begin{equation}
U= {\bar S (-\hat p^3 - \hat p^1) - (-\hat q^3 - \hat q^1) \over \bar
S (\hat
p^0 - \hat p^2) - (\hat q^0 - \hat q^2) } ={\bar S  p^3 - q_2 \over
\bar S  p^0
- p^1 }=z_3 \ .
\end{equation}

\section{String triality and STU black holes}

Our results allow for a comparison with the string triality picture as
described in \cite{Triality}.
There, a six-dimensional string, described by the low energy action
\begin{equation}
I_6=\frac{1}{2\kappa^2}\int d^6x \sqrt{-G}e^{-\Phi}\left[
R_G+G^{MN}\partial_M\Phi\partial_N\Phi
-\frac{1}{12}G^{MQ}G^{NR}G^{PS}H_{MNP}H_{QRS}\right]\
\label{d6}
\end{equation}
with $M,N=0,..,5$
was considered. This string might be a truncated version of a
heterotic
or a Type $II$ string. Upon toroidal compactification to $D=4$
one obtains a $N=2$ supergravity theory coupled to three vector
multiplets. The four-dimensional metric is related to the
six-dimensional one by
\begin{equation}
G_{MN}=\pmatrix{g_{\mu\nu}+A_\mu^mA_\nu^nG_{mn}&A_\mu^mG_{mn}\cr
A_\nu^nG_{mn}&G_{mn}}\ ,
\label{metricred}
\end{equation}
where the space-time indices are ${\mu,\nu}=0,1,2,3$ and the internal
indices
are $m,n=1,2$.  Two more  vectors arise from the reduction of the $B$
field.

One also finds six scalars, four of which are moduli of the
2-torus.  We parametrize
the internal metric and $2$-form as
\begin{equation}
G_{mn}=e^{\eta_3-\eta_2}\pmatrix{
e^{-2\eta_3}+a_3^2&-a_3\cr
-a_3&1} ,
\end{equation}
and
\begin{equation}
B_{mn}=a_2\,\epsilon_{mn}\ .
\end{equation}
$\eta_1$, the four-dimensional
dilaton, is given by
\begin{equation}
e^{-\eta_1}=e^{-\Phi}\sqrt{\det G_{mn}}=e^{-(\Phi+\eta_2)}\ .
\end{equation}
The sixth  scalar is the   axion $a_1$ which arises from dualization
of the three-form field strength in four dimensions.

The scalars are typically combined into three complex scalars, which
in
notation suitable for our previous sections are:
\begin{eqnarray}
z_1= S&=&a_1-ie^{-\eta_1} \ , \nonumber \\
z_2=T&=&a_2-ie^{-\eta_2}\ ,  \\
z_3=U&=&a_3-ie^{-\eta_3} \  \nonumber.
\end{eqnarray}
Here  $S$ is a dilaton-axion of the heterotic string, 
$T$ and $U$ are the K\"ahler form and the complex structure of the
torus.  These three scalars are obviously the ones considered so far
in this paper.  The four vectors are combined to a vector $A_\mu^a$
with $a=1,2,3,4$.  Details can be found, e.g.\ , in \cite{Triality}.
In fact, the electric and magnetic charges can be put together to an
$SP(8)$ vector as given in the earlier chapters.

The symmetry of this theory is $SL(2,Z)\times {O(2,2,Z)\over O(2)\times
O(2)}$.
The $SL(2,Z)$ component is the famous $S$-duality, a conjectured
non-perturbative symmetry of string theory. The second factor,
which is just a product of two $SL(2,Z)$ plus their exchange, is related to 
perturbative $T$-duality symmetry. In the following, $T$-duality will
denote the duality symmetry generated by the first $SL(2,Z)$, which
acts on the K\"ahler form, whereas the second one is called
$U$-duality
and acts on the complex structure $U$. All three symmetries act on
the scalars
by
\begin{equation}z_i\rightarrow \frac{a_i z_i+ b_i}{c_i z_i +d_i} \label{dualities}
\end{equation}
with $a_i d_i-b_i c_i=1$. The electric and
magnetic
charges transform as vectors under the three duality symmetries
(where the $SP(8)$ vector has to be converted into an $SL(2)^3$
vector \cite{Triality}).

This theory is precisely the one studied in \cite{ksw}.  In
\cite{Triality} it was found that the theory allows two (or five,
according to taste) dual descriptions where the roles of $S$, $T$ and
$U$ get interchanged. For example, $S$ is the dilaton/axion field for
the heterotic string, the K\"ahler form for the Type $IIA$ string and
the complex structure of the Type $IIB$ string. However, all those
theories were of the same type, in the sense that (at least in the
truncated versions considered here) two symmetries were perturbative
and one was non-perturbative.

This is easily seen by considering (for example) the
four-dimensional heterotic Lagrangian (in the absence of axionic
fields):
\begin{equation} \label{lagrang1}
{\cal L} = \sqrt{-g}\biggr[R-\frac{1}{2}\sum(\partial \eta_i)^2
-\frac{1}{4}\left(e^{-\eta_1-\eta_2-\eta_3}F^1 F^1+
             e^{-\eta_1-\eta_2+\eta_3} F^2 F^2+
             e^{-\eta_1+\eta_2+\eta_3} F^3 F^3+
            e^{-\eta_1+\eta_2-\eta_3} F^4 F^4\right)\biggr].
\end{equation}
Clearly, $\eta_2\rightarrow -\eta_2$ and $\eta_3\rightarrow -\eta_3$
(accompanied by an exchange of a few field strengths) are off-shell
symmetries, where as $\eta_1\rightarrow -\eta_1$ requires
dualizations
of field strengths.

How does the STU-model considered in this paper tie in with those
three (or six) string theories?  Obviously, it cannot correspond to
either of those, since it treats $S$, $T$ and $U$ on equal footing.
This is already clear from the prepotential, but also the action gives
some insights.  It can be obtained from
\begin{equation}{\cal L}_V= {\rm Im} {\cal N}_{\Lambda\Sigma}\ {\cal F}^\Lambda
 {\cal F}^\Sigma + {\rm Re}
 {\cal N}_{\Lambda\Sigma}\ {\cal F}^\Lambda \ {}^* {\cal F}^\Sigma.
\end{equation}
We find
\begin{equation}
F_{\Lambda\Sigma}=\pmatrix{2d_{ijk} z^iz^jz^k  &
     -3d_{mij}  z^iz^j  \cr
  -{3}d_{lij}  z^iz^j  & 6d_{mli}z^i  } \ ,
\end{equation}
from which one can deduce
\begin{equation}
{\cal N}_{\Lambda\Sigma}=\bar F_{\Lambda\Sigma}+2i \frac{({\rm
Im}F_{\Lambda\Omega})
({\rm Im}F_{\Pi\Sigma}) X^\Omega X^\Pi}
{({\rm Im }F_{\Omega \Pi}) X^\Omega X^\Pi} \label{N} \ ,
\end{equation}
where $X^\Lambda=(1,z^1,z^2,z^3)$. We do not try to express ${\cal N}$
in full generality, however we note that the lower three by three
matrix
${\cal N}_{ij}$ has the extremely simple form
\begin{equation}
{\cal N}_{ij}=\pmatrix{-ie^{+\eta_1-\eta_2-\eta_3}& a_3 &a_2 \cr
a_3 & -ie^{-\eta_1+\eta_2-\eta_3} & a_1 \cr a_2 & a_1 &
-ie^{-\eta_1-\eta_2+\eta_3}
}.
\end{equation}
The vector part of the  Lagrangian in the absence of any axion-like fields
is then given by
\begin{equation} \label{lagrang2}
{\cal L}_V=
-\left(e^{-\eta_1-\eta_2-\eta_3}{\cal F}^0{\cal F}^0+
             e^{+\eta_1-\eta_2-\eta_3}{\cal F}^1{\cal F}^1+
             e^{-\eta_1+\eta_2-\eta_3}{\cal F}^2{\cal F}^2+
            e^{-\eta_1-\eta_2+\eta_3}{\cal F}^3{\cal
F}^3\right)\biggr].
\end{equation}
Note that this Lagrangian has perfect exchange  $(1
\leftrightarrow2\leftrightarrow3)$ symmetry,
but it is not
invariant under any $\eta_i\rightarrow -\eta_i$ duality
transformation (accompanied by the appropriate exchange of vector
fields).
Hence, the theory has neither $S$,
$T$ nor $U$ duality (in the notation of \cite{Triality}) realized
off-shell!

We can also compare how $S$, $T$ and $U$ dualities (\ref{dualities})
are realized as $SP(8)$ matrices. For this we go first in a
basis $(S|TU)$ (heterotic string compactified on a two-torus).
This can be done by the symplectic rotation
\begin{equation}
 {\cal C}: \qquad (p^{\Lambda},q_{\Lambda}) \rightarrow
 (\tilde{p}^{\Lambda},\tilde{q}_{\Lambda})=
 (p^0,-q_1,p^2,p^3\,|\,q_0,p^1,q_2,q_3) \ .
\end{equation}
In this
new basis the $SL(2,Z)$ transformations are realized by a matrix
(\ref{spmatrix}) with \cite{wi/ka/lo/lu}
\begin{eqnarray}
 S \rightarrow \frac{aS +b}{c S +d}\ : \quad \mbox{by} \quad a \,A=
 d\, D=\, a\,d\,\delta_{ij} \ ; \quad
b\, B = c\, C  = b\,c\, \left (\matrix{
             0 & 1 & 0 & 0 \cr
             1 & 0 & 0 & 0 \cr
             0 & 0 & 0 & 1 \cr
             0 & 0 & 1 & 0 \cr
              }\right ) , \\
\nonumber \\
T \rightarrow \frac{aT +b}{c T +d} \ : \quad  \mbox{by} \quad B=C=0 \ , \quad A = (D^T)^{-1} = \left (\matrix{             d & 0 & c & 0 \cr
             0 & a & 0 &-b \cr
             b & 0 & a & 0 \cr
             0 &-c & 0 & d \cr
                }\right ), \\
\nonumber \\
U \rightarrow \frac{aU +b}{c U +d} \ : \quad \mbox{by} \quad B=C=0 \ , \quad A = (D^T)^{-1} = \left (\matrix{             d & 0 & 0 & c \cr
             0 & a &-b & 0 \cr
             0 &-c & d & 0 \cr
             b & 0 & 0 & a \cr
                }\right )  .
 \end{eqnarray}
As we can see, in this basis
only the $S$-duality is non-perturbative (exchange of electric
with magnetic charges) whereas the $T$-and $U$-duality acts diagonal,
i.e.\ exchange electric with electric and magnetic with magnetic
quantum numbers. Finally, to get the transformation for our
original charges we have to invert the transformation ${\cal C}$.
Combining all symplectic transformations we find for our
charges in the $STU$ basis the transformations
\begin{equation}
 \left(\matrix{  p^{\Lambda}\cr q_{\Lambda}\cr }\right)
\rightarrow  
\left(\matrix{ d \,p^0 + c\, p^1 \cr
  b\, p^0+a \,p^1 \cr d\, p^2 + c \,q_3 \cr d \,p^3 + c\, q_2 \cr
  a \,q_0 - b \,q_1 \cr -c \, q_0 + d\, q_1 \cr  b \, p^3 + a \,q_2 \cr
   b \,p^2 + a \,q_3 }\right )_{SL(2)_S}  ; \qquad \qquad
\left (\matrix{ d \,p^0 + c \,p^2 \cr d \,p^1+c\, q_3 \cr
 b \,p^0 + a\, p^2 \cr d \,p^3 + c \,q_1 \cr
 a \, q_0 -b \,q_2 \cr b \,p^3 +a \,q_1 \cr -c \,q_0 + d \,q_2 \cr
  b \,p^1 + a \,q_3 } \right)_{ SL(2)_T}  ; \qquad \qquad
\left(\matrix{ d \,p^0 + c \,p^3 \cr d \,p^1+c\, q_2 \cr
 d \,p^2 + c\, q_1 \cr b \,p^0 + a \,p^3 \cr
 a \, q_0 -b \,q_3 \cr b \,p^2 +a \,q_1 \cr  b \,q_1 + a \,q_2 \cr
 -c \,q_0 + d \,q_3  } \right)_{ SL(2)_U} .
\end{equation}

As expected, one finds that $W$ as given in eq.  (\ref{ww}) or the area of the horizon
(mass) is invariant under these transformations (or $(SL(2))^3$ symmetric).

Let us turn back to the Lagrangians (\ref{lagrang1}) and
(\ref{lagrang2}).  Both lagrangians are on-shell equivalent,
(\ref{lagrang1}) corresponds to the $(S|TU)$  basis whereas
(\ref{lagrang2}) is our $(STU)$ basis and the
transformation ${\cal C}$ maps both. This transformation is a
dualization of ${\cal F}_1$ and a renaming $({\cal F}_0, {\cal
F}_1, {\cal F}_2,{\cal F}_3)\rightarrow 1/2 (F_1,F_3,F_4,F_2)$.  Note,
that under dualization the pre-factor of ${\cal F}_1$ gets
inverted. This dualization makes all duality symmetries in the $(STU)$ basis
non-perturbative. The statement that in this symmetric basis
none of the dualities is perturbative is equivalent to the
statement that there is no basis in which all of dualities are
perturbative. The most we can achieve is to make only one
non-perturbative and two perturbative. We have considered the case
where the $S$-transformation is non-perturbative. On equal footing
we could take $T$ or $U$. These three possibilities fix then
the three underlying string theories (heterotic, type IIa or type IIb).
When all  three theories are symmetrised by going to the $STU$ basis,
immediately all dualities become non-perturbative.

These transformations can be nicely visualized in the form of a cube,
as it was done in \cite{Triality}. Figure 1 shows how the dualities
transform the field strengths into each other and their duals in the
$(STU)$ model.  Figure 1 and Figure 2 also illustrate the crucial
difference with the $(S|TU)$-, $(T|US)$- and $(U|ST)$-models of
\cite{Triality} (Figure 2). In the $STU$ theory, the fundamental field
strengths (or electric charges) are located around ${\cal F}^0$,
whereas in the $(S|TU)$, $(T|US)$ and $(U|ST)$ models the fundamental fields
were located on one side of the cube, allowing 2 dualities to be
perturbative.

The black holes with vanishing axions and finite scalars all have four
charges. These charges must be located on four corners of the cube
which are NOT connected by edges. Hence, the choices one has are
$q_0,p^1,p^2,p^3$ (with product of 4 charges $q_0p^1p^2p^3$ positive ) and
$p^0,q_1,q_2,q_3$ (with
 product of 4 charges $p^0 q_1 q_2 q_3$ negative), which is consistent with our
results. In the $S$-, $T$-, or $U$- string picture one always
needed two electric and two magnetic charges.

\begin{figure}
\centerline{\epsfbox{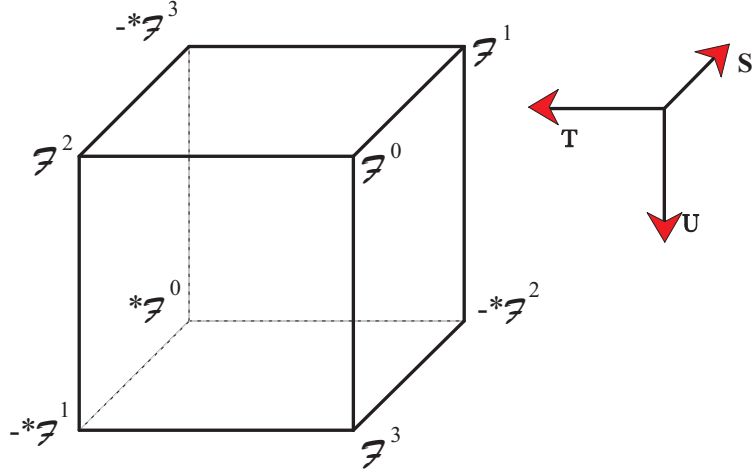}}
\medskip
\caption{Duality transformations in the STU-model. The fundamental
field strengths are not located on one side.}
\end{figure}

\begin{figure}
\centerline{\epsfbox{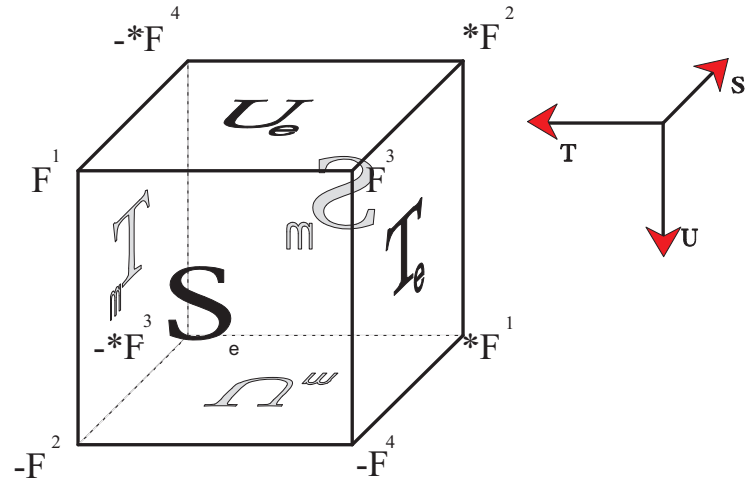}}
\medskip
\caption{Duality transformations in the S,T and U strings. The
fundamental
field strengths are located on one side and two duality symmetries
are
perturbative. The field strengths have different indices from Figure
one,
because the ones here are fundamental $S$-string fields.}
\end{figure}

\section{Outlook}

In conclusion, we have found a new type of d=4 supersymmetric black holes in the context of N=2 special geometry related to Calabi-Yau threefold. The main difference with the existing supersymmetric black holes is a completely {\it democratic treatment of all moduli of the theory}. This is due to our use of the version of special geometry with the 
prepotential $F=d_{ABC}\frac{X^AX^BX^C}{X^0}
$ \cite{CREM} where  $d_{ABC}$ are real symmetric constant tensors.  A  particular model of this type
with $F=STU$ gives no preference to any of the moduli and therefore none of them can  play a role of coupling constants. This makes the new $(STU)$  black holes different from stringy $(S|TU)$-, $(T|US)$- and $(U|ST)$  black holes \cite{ksw},
\cite{Triality}  where
one of the moduli ($S$ in heterotic case, $T$ in type IIa case and $U$ in type IIb case) does play the role of the coupling constant.

One may try to relate our new d=4 black holes to d=5 supersymmetric black holes described in \cite{FK}. The
area formula found there depends on symmetric tensor $d_{ABC}$ as follows.
\begin{equation}
Z_{\rm fix} =\sqrt {  \left (d^{AB}(q)\right)^{-1} q_A q_B} \  ,  \qquad A \sim \left [ \left (d^{AB}(q)\right)^{-1} q_A q_B \right ]^{3/4}  \ ,
\label{5darea}\end{equation}
 where $\left (d^{AB}(q)\right)^{-1}= \left (d^{AB} (t(z))|_{\partial_i Z  =0}\right)^{-1} $ and $(d^{AB})^{-1}$ is the inverse of $d_{ABC} t^C$.  Equation  (\ref{5darea}) applies in particular to eleven dimensional supergravity compactified on Calabi-Yau threefold. 

The new result found in this paper for four-dimensional black holes is the value of the moduli $t^A$  as the function of charges at the fixed point $\partial_i Z  =0$ for particular  example of  $d_{ABC}$. It can be used to find also the area of the five-dimensional black holes as the function of charges  for  this theory.

It remains to be seen if it is possible  to address  the issue of quantum corrections in string
theory using extreme black holes of classical moduli spaces as the starting
point. In this paper we have established  a duality relation between stringy $(S|TU)$-, $(T|US)$- and $(U|ST)$   black holes 
 and ``democratic"   $STU$ black holes.  Stringy black holes were known to be related to each other by the so-called triality in such a way that only one of $S,T,U$-dualities was non-perturbative \cite{Triality}.   The ``democratic" black holes
give us some new insights into the spectrum of states of the fundamental theory: all dualities there are non-perturbative.

\section{Acknowledgments}

This work was stimulated by the discussions with B. de Wit. 
The work of K.B. is supported by the DFG. He thanks the Physics
Department of the Stanford University for the hospitality and 
T.\ Mohaupt for many useful discussions. 
The work of R.K.,  J.R., M.S. and W.K.W  is supported by the NSF grant
PHY-9219345.
 The work of M.S. is supported by the Department of Energy under contract
DOE-DE-FG05-91ER40627.

\end{document}